\documentclass[twocolumn,epjc3]{svjour3}          

\RequirePackage[T1]{fontenc}

\smartqed  
\usepackage{amsmath}
\usepackage{empheq}
\RequirePackage{graphicx}
\RequirePackage{mathptmx}      
\RequirePackage{flushend}
\RequirePackage[numbers,sort&compress]{natbib}
\RequirePackage[colorlinks,citecolor=blue,urlcolor=blue,linkcolor=blue]{hyperref}

\journalname{Eur. Phys. J. C}

\begin{document}

\title{Reconstruction of a kinetic k--essence Lagrangian from a modified of dark energy equation of state}

\author{V\'ictor H. C\'ardenas\thanksref{e3,addr3}
\and Norman Cruz\thanksref{e1,addr1}
        \and
        Sebastian Mu\~noz\thanksref{e2,addr2} 
        \and
        J. R. Villanueva\thanksref{e4,addr3}
}
\thankstext{e3}{e-mail: victor.cardenas@uv.cl}
\thankstext{e1}{e-mail: norman.cruz@usach.cl}
\thankstext{e2}{e-mail: sebastian.munozc@usach.cl}
\thankstext{e4}{e-mail: jose.villanueva@uv.cl}

\institute{Instituto de F\'{\i}sica y Astronom\'ia, 
          Universidad de Valpara\'iso, Avenida Gran Breta\~na 1111, Casilla 5030, Valpara\'iso, Chile.\label{addr3} \and
          Departamento de F\'{\i}sica, Universidad de Santiago de Chile, Avenida Ecuador 3493, Casilla 307, Santiago, Chile.\label{addr1}
          \and
          Departamento de Matem\'atica y Ciencia de la
          Computaci\'on, Universidad de Santiago de Chile, Las Sophoras 173, Santiago, Chile.\label{addr2}
}
\maketitle

\abstract{
In this paper the Lagrangian density of a purely kinetic k--essence that models the behavior of dark energy described by four parameterized equations of state proposed by Cooray \& Huterer (Astrophys. J. {\bf 513} L95, 1999), Zhang \& Wu (Mod. Phys. Lett. A {\bf 27} 1250030, 2012), Linder (Phys. Rev. Lett. {\bf 90} 091301, 2003), Efstathiou (Mon. Not. Roy. Astron. Soc. {\bf 310} 842, 2000), and Feng \& Lu (J. Cosmol. Astropart. Phys.  {\bf 1111} 034, 2011) has been reconstructed. This reconstruction is performed using the method outlined by de Putter \& Linder (Astropart.\ Phys.\  {\bf 28} 263, 2007), which makes it possible to solve the equations that relate the Lagrangian density of the k--essence with the given equation of state (EoS) numerically. Finally, we discuss the observational constraints for the models based on 586 SNIa data points from the LOSS data set compiled by Ganeshalingam et al. (Mon. Not Roy. Astron. Soc. {\bf 433} 2240, 2013).

}

\tableofcontents

\section{INTRODUCTION}\label{intro}
It is well known that the analysis of the luminosity--redshift relation for distant type--SnIa supernovae suggested that the universe undergoes an accelerated expansion. This discovery about the accelerated expansion of the universe motivated scientists to propose models that would attribute these behaviors in the universe responsible for this expansion to an unknown energy component called dark energy, which is homogeneously distributed in the universe, and its pressure is negative. 
To explain the observed accelerated expansion of the universe, the simplest solution was Einstein's idea of vacuum energy, namely cosmological constant $\Lambda$. A small positive cosmological constant has been supported by a number of observations. Indeed, the cosmological constant is a perfect fit to the dark energy data. An important question related to the cosmological constant is the fact that the energy densities of dark energy and dark matter are now comparable. This is called the {\it coincidence problem}. From the cosmological constant problem we are motivated to find an alternative explanation for dark energy. By modifying the left--hand side of Einstein's equation, we get the modified gravity models. The idea of modified matter models is that the energy momentum tensor $T_{\mu \nu}$ contains exotic matter, which provides negative pressure, which consists of a canonical scalar field with a standard Lagrangian of the form $\mathit{L} = X - V(\phi)$. Modifying this canonical kinetic energy term in non-linear kinetic terms, the non-linear kinetic energy of the scalar field can drive the negative pressure without the help of a field potential. These models are called kinetic k--essence, which consists of a scalar field  described by a Lagrangian of the form $\mathit{L} = F (X, \phi)$ (see \cite{Callan,gibbons,Gibbons:2001gy,Sen:2002in,Sen:2002nu}). This generalization of the canonical scalar field models can give rise to new dynamics not possible in quintessence. The non-linear kinetic energy terms are thought to be small and usually ignored because the Hubble expansion damps the kinetic energy density over time. But what happens if there is a dynamical attractor solution which forces the non-linear terms to remain non-negligible? This is the main idea of the k--essence. In the context of cosmology, k--essence was first studied as a model for inflation (k--inflation) \cite{Armen}, and due to its dynamics, increasing interest has been devoted to it in cosmological investigations. In this paper, we restrict ourselves to a particular case of kinetic k--essence, which is to take only kinetic terms, i.e., assume that the Lagrangian is of the form $\mathit{L} = F (X)$  \cite{Sche}. Such scalar fields can be interpreted as barotropic perfect fluids. The classification of scalar field models that satisfy the condition of barotropic perfect fluid are detailed in \cite{Arro}. In \cite{Ro} the physical stability restrictions for $C_s ^2$ is discussed, as well as the dynamics of four types of dark energy (EoS) models is analyzed for the behavior of k--essence and quintessence. In \cite{Yan} the behavior of a model where the equation of state is a power law of the kinetic energy, $X$ ($\omega=\omega_0\, X^{\alpha}$), and the conditions needed for to have accelerated phases. Variable EoS were considered in \cite{Co}, where $\omega(z)$ was written as $\omega_0 + z($d$w/$d$z)_0$, and by means of gravitational lensing, the expected accuracy to which  today's equation of state  $\omega_0$ and its rate of change $($d$w/$d$z)_0$ can be simultaneously constrained was studied. In \cite{Zha}, from the joint analysis of four observations (SNe + BAO + CMB+ H0), constraints for the time-varying dark energy EoS $\omega(z)$ were obtained, where $\omega(z)$ was parameterized via two parameters and some of them depend on a given extra parameter. A classification into two types of models for the suitable $\omega(z)$ was made according to the boundary behavior and the local extreme point of the EoS $\omega(z)$.

As we have mentioned, this paper is devoted to presenting a method to reconstruct the Lagrangian of kinetic k--essence, using as initial information a specific parametrization of the EoS, the dark energy of the form $\omega = \omega (a)$. To do so, this paper is organized as follows: In Section \ref{kke} we review the main aspects of the kinetic k--essence model as a dark energy model. In Section \ref{npkkel}  we develop the parameterizations that justify the model presented by Efstathiou, which modify the original one presented in \cite{Feng}. Also, the evolution of these parameterizations, either for the Lagrangian and for $X$ as a function of the scale factor, are found. Then, by defining the general relation $\frac{\textrm{d}a}{\textrm{d}X}=\mathcal{H}(a,X)$, the kinetic k--essence Lagrangian is constructed follow by an analysis of the behavior of the generating function $F$. In Sec. \ref{obsconst} we use the parameterizations defined above to obtain the best reconstruction of $F(X)$ based on observational data, consisting of 586 SNIa data points from the LOSS data set compiled by Ganeshalingam et al. \cite{Ganeshalingam:2013mia}. Finally, in Sec. \ref{concl} we ended the work with final remarks and conclusions.

\section{KINETIC K--ESSENCE}\label{kke}

Our main goal is to investigate a dark energy model described by an effective minimally coupled scalar field $\phi$ with a non-canonical kinetic term, specifically to the so--called {\it puerely kinetic k--essence} model, in which the Lagrangian density $\mathcal{L}=-F( X)$ depends only on the kinetic terms $X=\frac{1}{2}\partial _{\mu}\phi\partial^{\mu}\phi$. This model can be obtained from the action for a k--essence field minimally coupled to gravity, given by

\begin{equation}
S=-\int {\rm d}^4 x\sqrt{-g}\left( F(X)+\frac{1}{2}R \right)\label{action2}
\end{equation}
in the background of an homogeneous and isotropic flat universe described by  Friedmann--Lema\^itre--Robertson--Walker metric:
\begin{equation}
\label{flrwM} {\rm d}s^2={\rm d}t^2-a(t)^2\left[{\rm d}r^2+r^2\left({\rm d}\theta^2+\sin^2 \theta\,{\rm d}\phi^2\right)\right].
\end{equation}
Unless otherwise stated, we consider $\phi$ to be smooth on the scales of interest so that $X=\frac{1}{2}\dot{\phi}^2$. The energy--momentum tensor derived from the action (\ref{action2}) allows us to identify the energy density $\rho$ and pressure $p$ with the following expressions  

\begin{equation}\label{Densidad}
\rho =F -2X\,F_X ,
\end{equation}  
and 

\begin{equation}
p = -F\label{Presion},
\end{equation} 
where $F_X\equiv {\rm d}F/{\rm d}X$. Assuming that the energy density is a positive quantity, we demand that $F -2X\,F_X\geq0$. Therefore, with $F>0$, the parameter of the equation of state becomes
\begin{equation}
\omega = \frac{p}{\rho} = \frac{F}{2X\,F_X -F}\label{EoS}.
\end{equation} 
An important result relating $F$, $X$ and the scale factor $a$ is obtained when the scalar potential becomes constant, so the Eqs. of motion yield to the following expression \cite{Sche}:
\begin{equation}
X\,F_X ^{2} = \kappa \,a^{-6},\label{Rest1}
\end{equation} where $\kappa$ is a constant of integration. Thus, given any form of $F(X)$, the above equation gives the evolution of $X$ as a function of the scale factor $a$, and then, by substitution of this solution into Eqs. (\ref{Densidad})-(\ref{Presion}), we can obtain the evolution of all of the physical quantities of interest.

The square of the adiabatic speed of sound is obtained from the relation \cite{Garr}:

\begin{equation}
c^{2}_s = \frac{{\rm d}p/{\rm d}X}{{\rm d}\rho /{\rm d}X}=\frac{F_X}{F_X+2X\,F_{XX}}=\frac{3\omega (1+\omega)-\omega'}{3(1+\omega)}\label{Vel.sonido},
\end{equation}
where $F_{XX}\equiv {\rm d}^2F/{\rm d}X^2$, and a prime denotes differentiation with respect to $\ln a$. Notice that Eq. (\ref{Vel.sonido}) imposes two conditions: the first one, since the speed of sound is a real quantity, we demand that $3\omega (1+\omega)-\omega'\geq0$ \cite{Ro}. The second one comes from the fact that the speed of sound is slower than the speed of light (which is equal to 1 in this work), so we must demand that $3\omega (1+\omega)-\omega'\leq 3(1+\omega)$.
Finally, notice that, in general, the adiabatic speed of sound $c_s$ does not coincide with the speed of propagation of scalar perturbations. However, in the kinetic k--essence scenario this statement does not work and these two speeds match, which implies that it is possible describe a scalar field by a perfect fluid and vice versa \cite{Arro}.

Now, by combining Eqs. (\ref{Presion}) and (\ref{EoS}), it is possible to show that
\begin{equation}
X^2\,F_X ^2 = \left(\frac{1 +\omega}{2\omega} \right)^2\, p^2,
\end{equation} 
and then using the following identities
\begin{eqnarray}
\label{eqfr1}
\frac{1+\omega}{\omega}&=&\frac{p+\rho}{p},\\\label{eqfr2}
p'&=&[\omega'-3\omega(1+\omega)]\,\rho
\end{eqnarray}
we obtain the following equations for $X$ and $F$ in terms of scale factor:
\begin{equation}
X'= -6\,c^{2}_s\, X\label{EcX},
\end{equation}
and
\begin{equation}
F' =-3\left(\frac{1+\omega}{\omega}\right)\,c_s^2\,F\label{EcF}.
\end{equation}
Since the EoS parameter $\omega$ depends on the scale factor $a$, we can to introduce here the model to consider. Finally, by solving Eqs. (\ref{EcX}) and (\ref{EcF}),  we can write

\begin{equation}
F(X) = C\frac{\omega(a(X))}{a^3 (X) (1+\omega(a(X)))} \sqrt{X}\label{Lagragiano}
\end{equation}

Obviously, the main difficulty is to invert the expression $X(a)$, so we will perform a numerical approach to obtain solutions. In the next section we begin defining an expression  $\frac{d a}{dX}=\mathcal{H}(a,X)$ which, together with Eq.  (\ref{Lagragiano}) and the model proposed for the EoS, allows us to determine the phenomenological behavior of $F$ as a function of $X$.

\section{NEW PARAMETRIZATIONS AND KINETIC K--ESSENCE LAGRANGIANS}\label{npkkel}

As we have mentioned, the main goal of this research is to obtain a parametric expression relating $F$ and $X$ by means of the parametrization of the scale factor $a$, i.e., assuming that we know $F(a)$ and $X(a)$. In what follows, we display some proposed parameterizations \cite{Co}$-$\cite{Ef} for the EoS of dark energy, and their main drawbacks. The explicit expressions are the following:

\begin{subequations}
    \begin{empheq}[left={\frac{\omega-\omega_0}{\omega_a}=\empheqlbrace\,}]{align}
      & z=\frac{1-a}{a},\,\,\,\,\,\,\,\,\,\,\,\,\,\,\,\,\,\,\,\,\,\,\,\,\,\,\,\,\,\,\,\,\,\,\,\,\,\,\,\,\,\,\,\,\,\,\,\,\,\,\,\,\,\,\,\,\,\,\,\,\,\,\,\, \textrm{\cite{Co,Li}}\label{EoSvarias1} \\
      &  \frac{z}{1+z} = 1-a,  \,\,\,\,\,\,\,\,\,\,\,\,\,\,\,\,\,\,\,\,\,\,\,\,\,\,\,\,\,\,\,\,\,\,\,\,\,\,\,\,\,\,\,\,\,\,\,\,\,\,\,  \textrm{\cite{Zha,Li}} \label{EoSvarias2} \\
      & \ln (1+z)     = -\ln a, \,\,\,\,\,\,\,\,\,\,\,\,\,\,\,\,\,\,\,\,\,\,\,\,\,\,\,\,\,\,\,\,\,\,\,\,\,\,\,\,\,\,\,\,\,\,\,\,\, \textrm{\cite{Ef}} \label{EoSvarias3} \\
      & \ln \left(1+\frac{z}{1+z} \right)=  \ln \left( 2-a \right),\,\,\,\,\,\,\,\,\,\,\, \textrm{\cite{Feng}} \label{EoSvarias4}
    \end{empheq}
\end{subequations}
where $\omega_0$ is the present value of the parameter $\omega$,  $\omega_a$ should be adjusted to the observational data. In each parametrization  $\omega_0$ represents the value of $\omega$ at the present time and if $\omega_0=-1$ and $\omega_a = 0$ they lead to the classical model $\Lambda$CDM. The first parametrization represents a good fit for small $z$, but has a serious problem explaining the observations for large $z$. ($z > 1$). The second shows good behavior when $z<1$ and $z>1$. This is because for $z<1$, $\omega$ has an approximately linear behavior while for $z\gg 1$, $\omega$ is bounded. In the parameterization proposed in \cite{Ef},  when  $z \rightarrow \infty$, $w(z)$ becomes infinite, so this parameterization can only describe the behavior of dark energy when $z$ is not very large. In order to avoid this problem a  modification was introduced in \cite{Feng} assuming the form $w(z) = w_0 + w_a \ln (1 + z/1+z)$. For future evolution, when  $z \rightarrow −1$,  $\left| w(z) \right|$ will grow rapidly and diverge, which is a nonphysical behavior of this EoS. 

We show below explicitly the expression for $F(a)$ and $X(a)$ for the dark energy models proposed in \cite{Co,Zha,Li,Ef}, using Eq. (\ref{EcX}) and Eq. (\ref{EcF}); however, the reconstruction of the corresponding  kinetic k--essence model, $F(X)$, will be done only for the new parameterization proposed by Feng \& Lu in \cite{Ef}, which represents an improvement in the behavior of the dark energy EoS for a wide range of the redshift.

Using the parametrizations proposed in  \cite{Co,Zha,Li,Ef} in Eq. (\ref{EcX}), we can solve for $X(a)$, yielding

\begin{subequations}
    \begin{empheq}[left={\frac{X(a)}{X_0}=\empheqlbrace\,}]{align}
      & \left(\frac{\omega_0}{a} + 1+\omega_0 - \omega_a \right)^{2} a^{6(\omega_a - \omega_0)} e^{\left(\frac{6\omega _a}{a} \right)}.\label{sol1} \\
      &  (1+\omega_0+\omega_a -\omega_a a)^2 a^{-6(\omega_0 + \omega_a)} e^{(6\omega_a a)}\label{sol2} \\
      & (1+ \omega_0-\omega_a \ln a)^{2} a^{-6\omega_0} e^{(3\omega_a \ln ^2 a)} \label{sol3}
    \end{empheq}
\end{subequations}

Then using Eq. (\ref{EcF}), the corresponding Lagrangian kinetic k--essence densities as a function of the scale factor are obtained:

\begin{subequations}
    \begin{empheq}[left={\frac{F(a)}{F_0}=\empheqlbrace\,}]{align}
      & \frac{\left( \omega_a+(\omega_0-\omega_a )a \right)^{-\frac{3}{\omega_a}}\exp \left(\frac{3\omega_a}{a} \right) }{ a^{3(\omega_0 +1 - \omega_a-\frac{1}{\omega_a})}}.\label{func1} \\
      &  \frac{\left(\omega_0+\omega_a -\omega_a a\right) \exp (3\omega_a a) }{a^{3(1+\omega_0 + \omega_a)} }\label{func2} \\
      & \frac{\left(\omega_0-\omega_a \ln a\right) \exp \left(\frac{3}{2}\omega_a \ln ^2 a\right)}{a^{3(1+\omega_0)} } \label{func3}
    \end{empheq}
\end{subequations}
Using these equations we can reconstruct $F=F(X)$ by inverting the expression $X(a)$, but we need $\frac{\textrm{d} X}{\textrm{d} a} \neq 0$ to guarantee the existence of the inverse. In what follows, we will do an explicit reconstruction of $X$ for the  parameterizations proposed in \cite{Feng} and given by Eq. (\ref{EoSvarias4}).
This ansatz has great interest in their behavior as a dark energy model, because it combines the advantages of the previous parameterizations. Also, this model has a future singularity at $z=-\frac{1}{2}$, which is a value bounded by the big bang. Depending on the dynamics, it may be adjusted to show a phantom-type  behavior for a given redshift; however, in \cite{Durrer} it is discussed that the k--essence models entail problems in the case where the values of the adiabatic sound speed across this barrier. This detail is dependens on the chosen parametrization and the action taken initially. 

Imposing the Eq. (\ref{EcF}), we cannot explicitly obtain  $F(X)$ for this particular model. However, if we manipulate this expression and solve an equation of the form $\frac{\textrm{d}a}{\textrm{d}X}=\mathcal{H}(a,X)$ numerically, and we use Eq. (\ref{Lagragiano}), we can then obtain approximate values for $ F (X) $ with $ X> 0 $.
We rewrite Eq. (\ref{EcX}) as follows

\begin{equation}
\frac{\textrm{d}a}{\textrm{d}X}=\mathcal{H}(a,X)=-\frac{a}{6 C_s ^2 (a)X}\label{Ec.a(X)},
\end{equation}
where the adiabatic sound speed squared $ C_s ^ 2 $ for the model given by Eq. (\ref{EoSvarias4})  is given by

\begin{equation}
C_s ^2 = \omega_0 + \omega_a \ln (2-a) + \frac{\omega_a a}{3(2-a)(1+\omega_0 +\omega_a \ln (2-a))} \label{SonidoEf}.\\
\end{equation}
Then, introducing Eq. (\ref{SonidoEf}) into Eq. (\ref{Ec.a(X)}) we obtain

\begin{eqnarray}\nonumber
&&\frac{da}{dX}=-\frac{a}{2X}\times \\\label{ax}
&&\left(\frac{(1+ \omega_0 + \omega_a \ln (2-a))}{(\omega_0 + \omega_a \ln (2-a))(1+ \omega_0 + \omega_a \ln (2-a))-\frac{\omega_a a}{a-2}}\right)\label{Ec.a(X)2}.\\\nonumber
\end{eqnarray}

In \cite{Feng} it was determined that the values which fit the observational data are $ \omega_0 = -1.0537 $ and $\omega _a = 0.2738$. Using  these values, we plotted the behavior of $a=a(X)$ (see Fig \ref{fig:fig1}). This numerical solution was obtained using the Runge-Kutta method (see \cite{Dormand}) and is the first step in reconstructing process of the kinetic k--essence Lagrangian. In Fig. \ref{fig2} the behavior of $F(X)$ as a function of $X$ is displayed, where the initial condition chosen was $a(X=1)=1$.

It is straightforward to see from Fig. \ref{fig2} that the reconstructed k--essence Lagrangian is a decreasing function of the kinetic energy, $X$, of the field.  Note that the reconstruction in $X=0$ is not possible due the indetermination in the values of $F_X$, which can be appreciated in Eq. (\ref{Rest1}). From direct inspection of Eq. (\ref{EoS}) we can conclude that\\

\begin{center}
	\begin{tabular}{ccc}
		$\omega +1 > 0 $& $\Rightarrow$ & $F_X < 0$ \\\\
		$\omega +1 < 0 $& $\Rightarrow$ & $F_X > 0$ \\\\
	\end{tabular} 
\end{center}

The above conditions imply that, if $F >0$, phantom behaviors can be obtained from kinetic k--essence Lagrangians which are increasing functions of $X$, whereas quintessence behaviors ($\omega > -1$) are obtained from decreasing functions.   On the other hand, from the conditions imposed on $C_s ^2$ \cite{Ro}, we obtain

\begin{center}
	\begin{tabular}{ccc}
		$\omega  > -1 $& : & $F_{XX} > 0$ \\\\
		$\omega  < -1 $& : & $F_{XX} < 0$ \\\\
	\end{tabular} 
\end{center}

So, Lagrangian leading to an EoS in the range of quintessence will be decreasing functions of $X$ but with a positive second derivative, whereas for those which represent phantom EoS, the Lagrangian is an increasing function of $X$, but with a negative second derivative or, in other words, it will be a concave function.  

\section{OBSERVATIONAL CONSTRAINTS}\label{obsconst}

In this section we use the parameterizations (\ref{EoSvarias1}-\ref{EoSvarias4}) to obtain the best reconstruction of $F(X)$ based on
observational data. In particular, we make use of the LOSS data set
\cite{Ganeshalingam:2013mia} consisting in 586 SNIa data points. The
SNIa data enable us to constrain the luminosity distance
$d_L(z)=(1+z)r(z)$, where $r(z)$ is the comoving distance. We fit
the SNIa with the cosmological model by minimizing the $\chi^2$
value defined by
\begin{equation}
\chi_{SNIa}^2=\sum_{i=1}^{586}\frac{[\mu(z_i)-\mu_{obs}(z_i)]^2}{\sigma_{\mu
i}^2},
\end{equation}
where  $\mu(z)\equiv 5\log_{10}[d_L(z)/\texttt{Mpc}]+25$ is the
theoretical value of the distance modulus, and $\mu_{obs}$ is the
corresponding observed one.

Actually, there is no need to use (\ref{EcX}) and (\ref{EcF}) to
obtain $X(a)$ and $F(a)$ for a given $\omega(a)$. In fact, from
(\ref{Presion}) we know that
\begin{equation}\label{ffa}
F(a) = \omega(a) \rho(a),
\end{equation}
and also from (\ref{Rest1}) we know that
\begin{equation}\label{kka}
4kX(a) = a^{-6}(1+\omega(a))^2\rho(a)^2.
\end{equation}
So, all we need is the expression for $\rho(a)$. From the mass
conservation equation we can write
\begin{equation}
\rho(z) = \rho_0 (1+z)^3 \exp \left(3 \int_0^{z}
\frac{\omega(x)}{1+x}dx \right),\label{rhoz}
\end{equation}
where we have introduced the redshift $z$ defined through the
relation $a=(1+z)^{-1}$.

\subsection{Model A}

Let us start with a first case, the one we called Model A, that is
defined through its EoS parameter given by $w(z)=w_0 + w_a z$. Using Eq.
(\ref{rhoz}) we obtain
\begin{equation}
\rho(z) = \rho_0 (1+z)^{3(1+w_0-w_a)} e^{3 w_a z}.
\end{equation}
and from Eq. (\ref{ffa}) we get
\begin{equation}\label{fasz}
F(z) = \rho_0 (w_0 + w_a z)(1+z)^{3(1+w_0-w_a)} e^{3 w_a z}.
\end{equation}
Therefore, by using Eq. (\ref{kka}) we obtain
\begin{equation}\label{xasz}
4k X(z) = \rho_0^2 ( 1 + w_0 + w_a z)^2 (1+z)^{6(w_0-w_a)} e^{6 w_a
z}.
\end{equation}
With these expressions we can plot $F(X)$ parametrically using the
scale factor (or the redshift) as the parameter. The values for
$w_0$ and $w_a$ in the plots are obtained from a test of the model
against observational data, in this case a Type Ia supernova.
\begin{figure}[h!]
\centering
  \begin{tabular}{@{}cc@{}}
    \includegraphics[width=.43\textwidth]{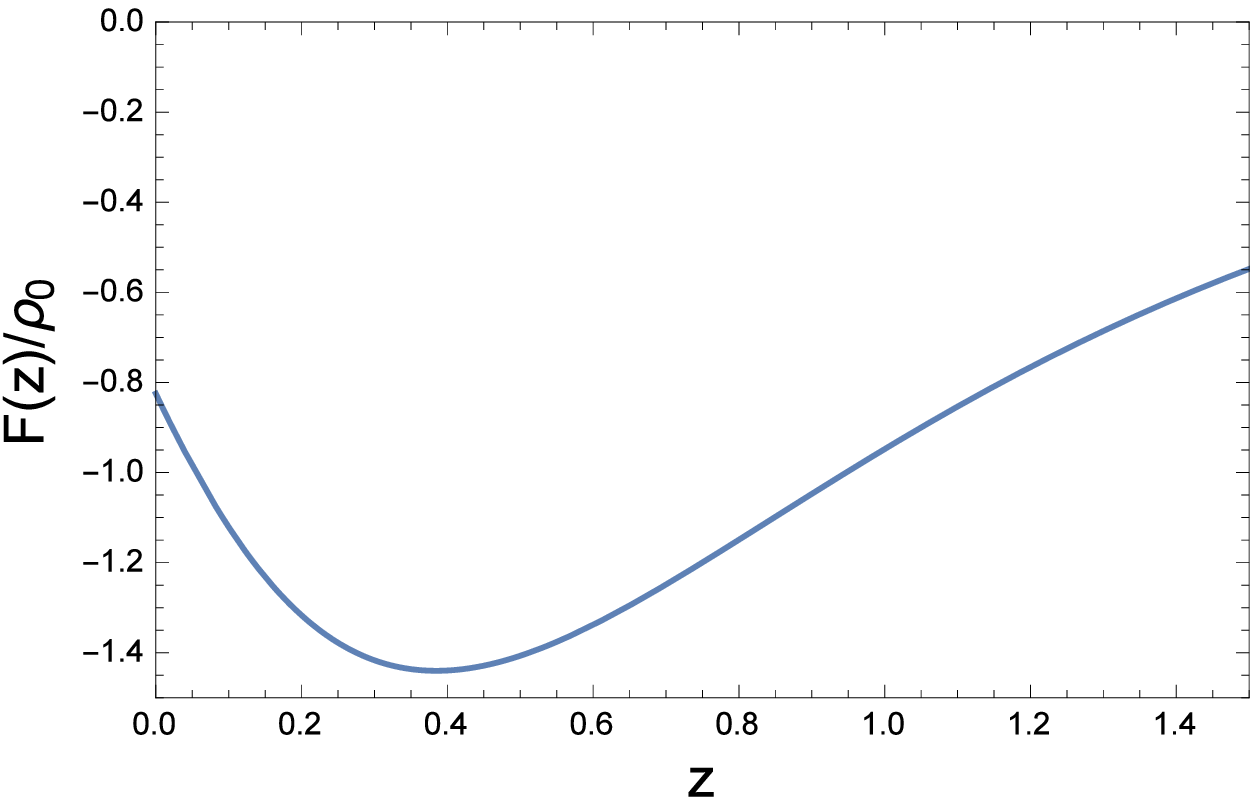} \\
    \includegraphics[width=.43\textwidth]{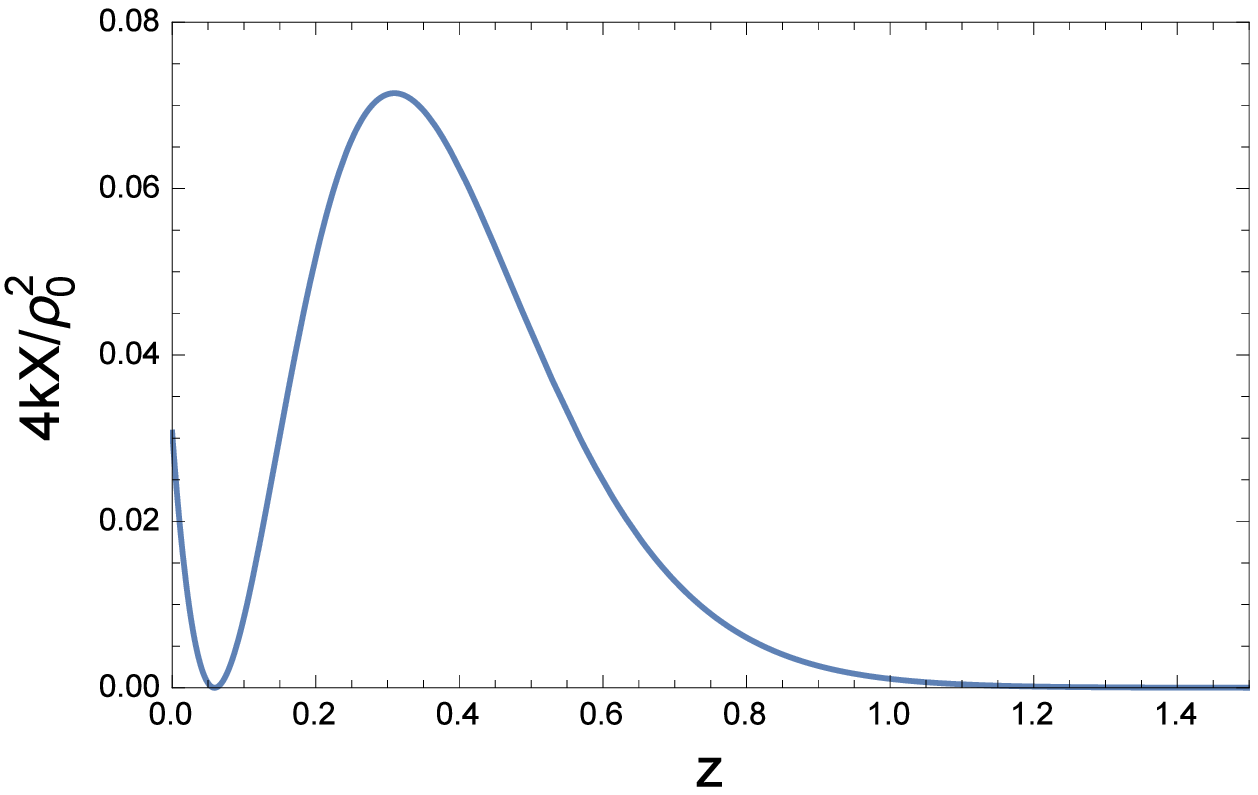}
  \end{tabular}
  \caption{Here we display $F$ and $X$ as a function of redshift
  based on Eqs. (\ref{fasz}) and (\ref{xasz}) for model A
  characterized by the parametrization $w_0+w_a z$.
  }\label{fig:fig1}
\end{figure}
The best fit to SNIa data gives $\Omega_m = 0.36 \pm 0.06$, $w_0 =
-0.8 \pm 0.2$, and $w_a = -3 \pm 2$, with $\chi^2_{red} = 0.984$. In
the first panel of Fig. \ref{fig:fig1} we display $F$ as a
function of redshift $z$, and in the other $X$ as a function of $z$.
Combining these two functions, we plot in Fig. \ref{fig2} the
reconstructed $F(X)$ using the best fit values of the parameters.

\begin{figure}[h!]
\centering
  \begin{tabular}{@{}c@{}}
    \includegraphics[width=.43\textwidth]{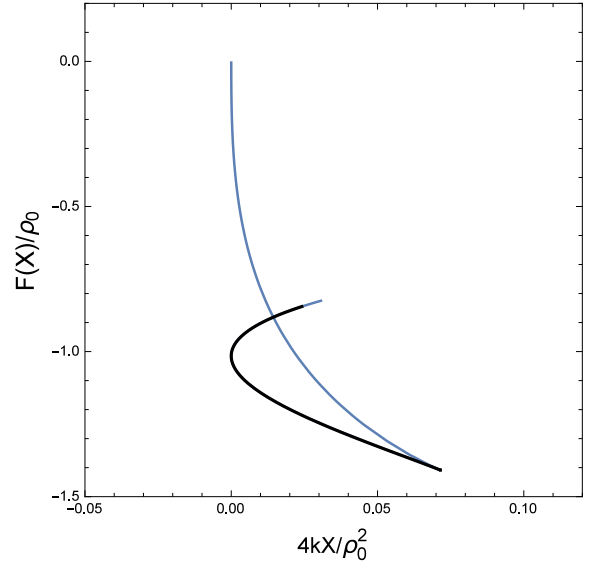}
  \end{tabular}
  \caption{Here the reconstructed $F(X)$ is shown based on the parametrization $w_0+w_1 z$ with the best fit values of $w_0, w_a$ obtained from SNIa data
  points. The black line indicate the segment in which $c_s^2>0$.
  }\label{fig2}
\end{figure}
A stability criterion demands that the sound speed be positive
$c_s^2>0$, which translates into a restriction of the values of the
EoS parameter $w(z)$. Using Eq. (\ref{Vel.sonido}) we notice
that $c_s^2>0$ means $X'<0$, or in terms of a derivative respect to
redshift, that $dX/dz>0$. From inspection in Fig. \ref{fig:fig1}
for model A, this only happens in the range $z\in [0.0591,0.3095]$.
In Fig. \ref{fig2} we represent such a region, drawing it as a
black thick line. Notice that $F(X)$ is double-valued and always
negative.

The same procedure is performed for the other three
parameterizations:
\begin{eqnarray} \nonumber
  \texttt{Model B} &  w(z)= w_0 + w_a z/(1+z)   \\ \label{models}
  \texttt{Model C} &  w(z)= w_0 + w_a \ln(1+z)  \\  \nonumber
  \texttt{Model D} &  w(z)= w_0 + w_a \ln \left(1+ \frac{z}{1+z}\right)  \\ \nonumber
\end{eqnarray}
and the results are displayed in Table \ref{tI}.
\begin{table}[h!]
\begin{center}
\begin{tabular}{c|c|c|c}
  \hline
   & $\Omega_m$ & $\omega_0 $ & $\omega_a$ \\
   \hline
Model A  & $0.36 \pm 0.06$ & $-0.8 \pm 0.2$  & $-3 \pm 2$ \\
Model B  & $0.35 \pm 0.06$ & $-0.8 \pm 0.2$  & $-4 \pm 3$ \\
Model C  & $0.35 \pm 0.06$ & $-0.8 \pm 0.2$  & $-3 \pm 3$ \\
Model D  & $0.34 \pm 0.06$ & $-0.7 \pm 0.2$  & $-4 \pm 3$ \\
\hline
\end{tabular}
\caption{Summary of the result of the reconstruction for each model.}\label{tI}
\end{center}
%
\end{table}

\subsection{Model B}

\begin{figure}[h!]
\centering
  \begin{tabular}{@{}cc@{}}
    \includegraphics[width=.43\textwidth]{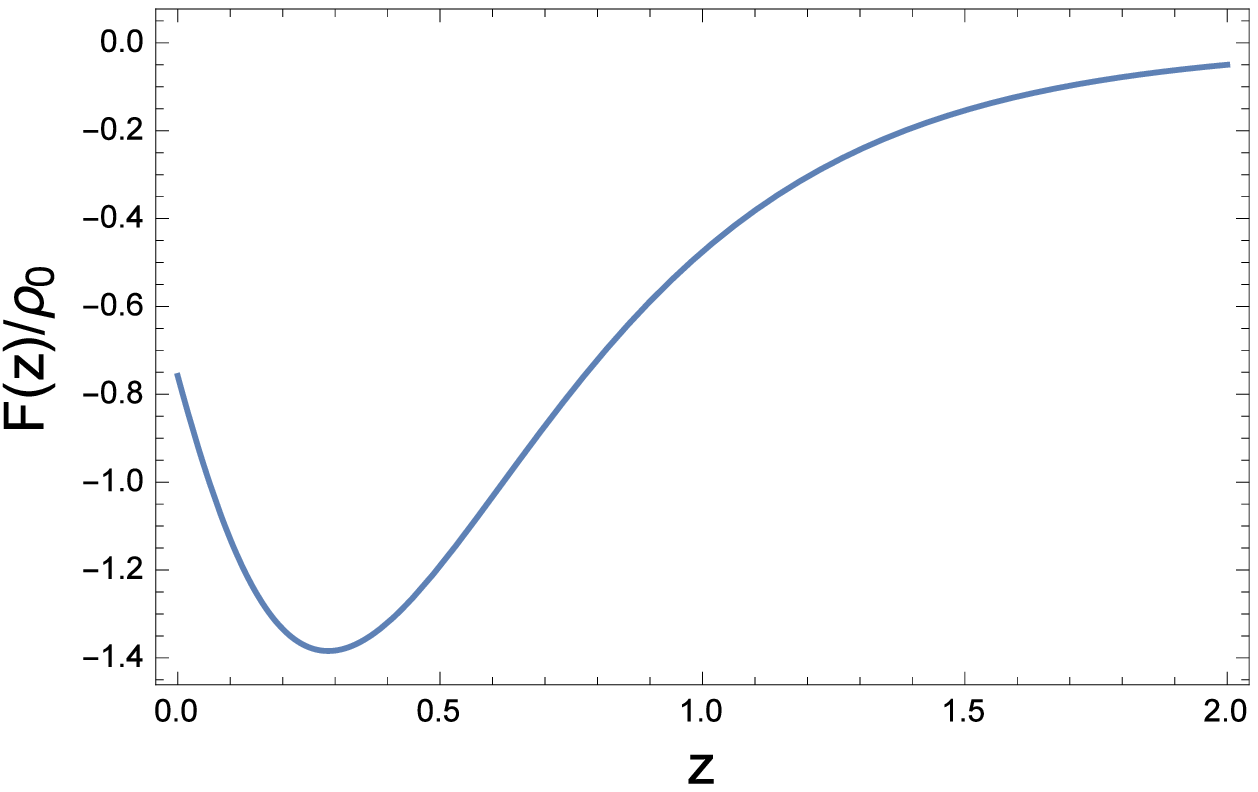} \\
    \includegraphics[width=.43\textwidth]{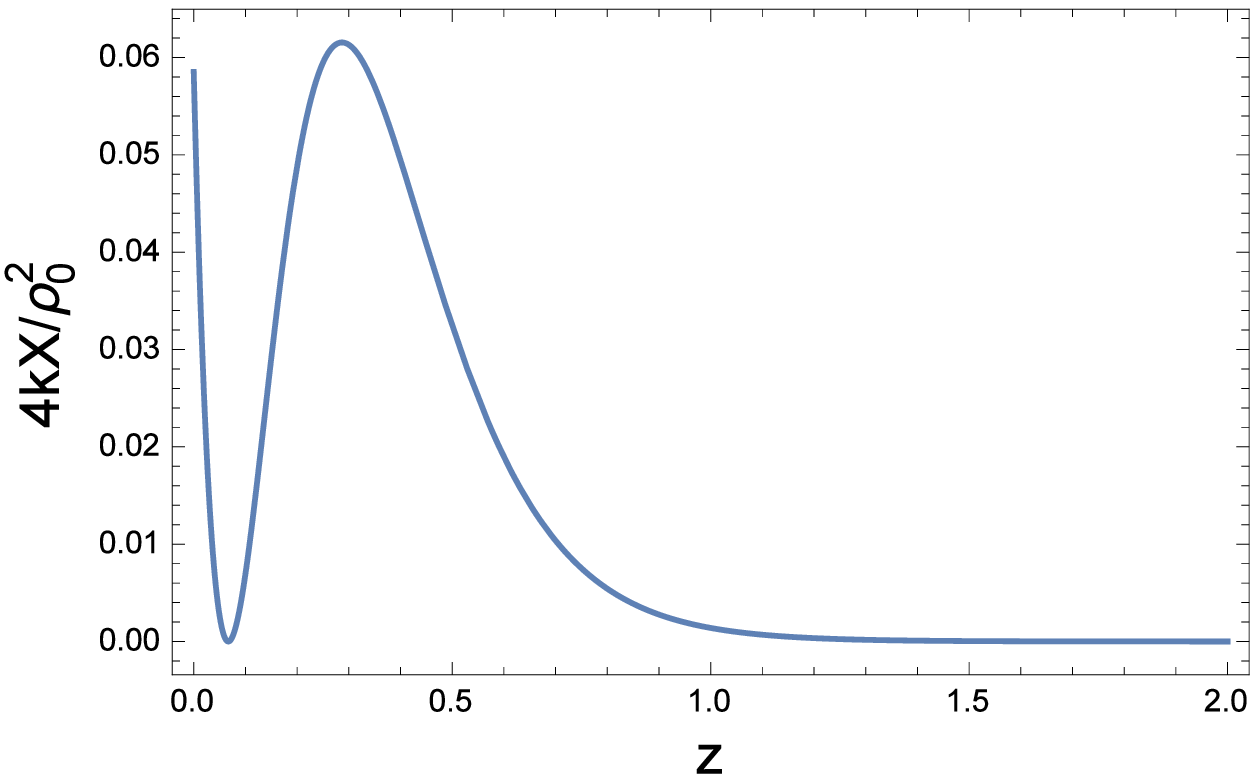}
  \end{tabular}
  \caption{Here we display $F$ and $X$ as a function of redshift
  based on Eqs. (\ref{fbsz}) and (\ref{xbsz}) for model B.
  }\label{fig3}
\end{figure}

\begin{figure}[h!]
\centering
  \begin{tabular}{@{}c@{}}
    \includegraphics[width=.43\textwidth]{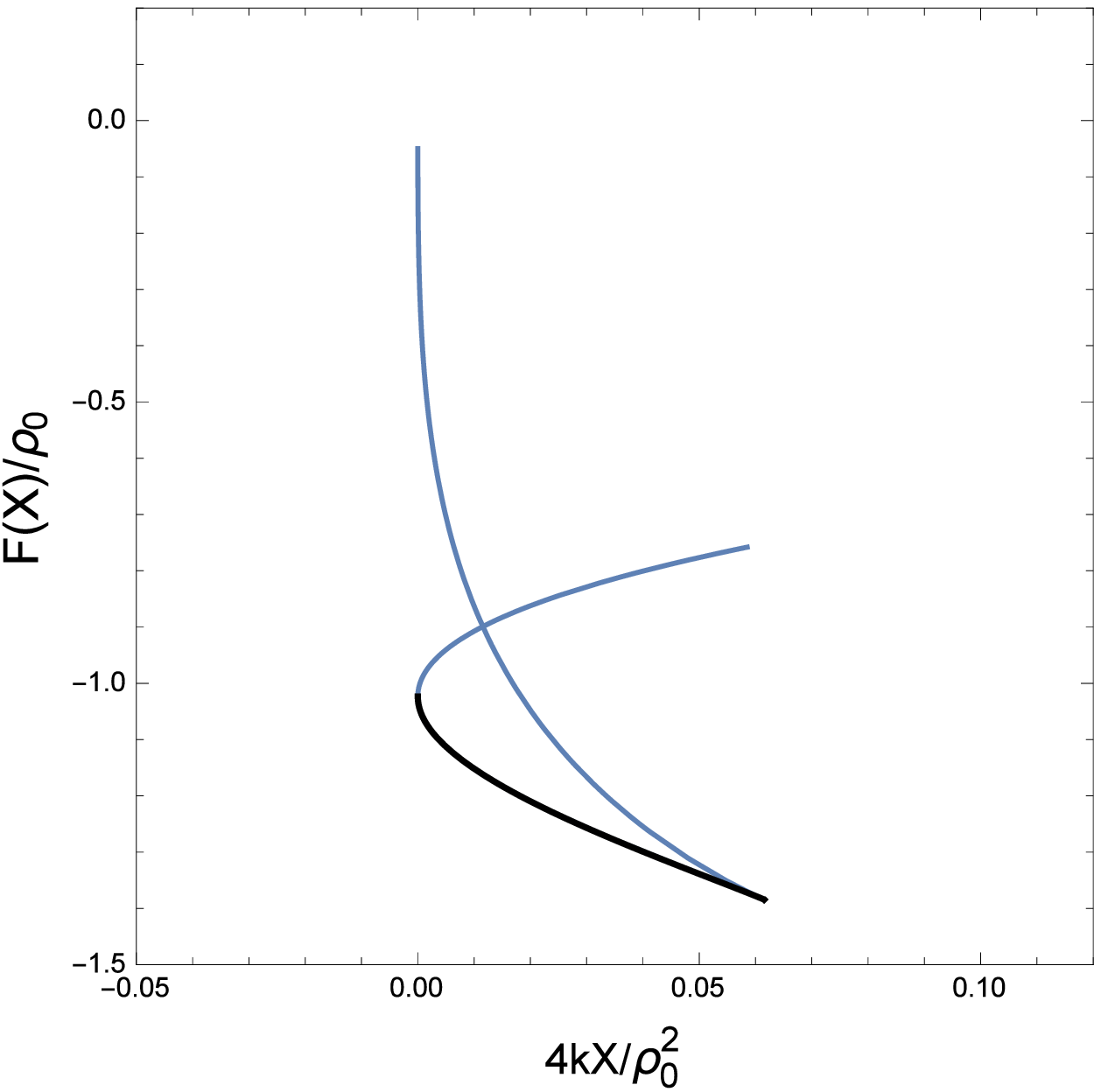}
  \end{tabular}
  \caption{Here the reconstructed $F(X)$ is shown based on the parametrization $w_0+\frac{w_az}{1+z}$ with the best fit values of $w_0, w_a$ obtained from SNIa data points. The black line indicates the segment in which $c_s^2>0$.
  }\label{fig:fig0b}
\end{figure}

For model B we have the Chevallier--Polarski--Linder (CPL) parameterization $w(z)=w_0 + w_a z/(1+z)$. From Eq. (\ref{rhoz}) we obtain
\begin{equation}
\rho(z) = \rho_0 (1+z)^{3(1+w_0+w_a)} e^{-3 \frac{w_a z}{1+z}}.
\end{equation}
and from Eq. (\ref{ffa}) we get
\begin{equation}\label{fbsz}
F(z) = \rho_0 (w_0 + \frac{w_a z}{1+z})(1+z)^{3(1+w_0+w_a)} e^{-3\frac{w_a z}{1+z}}.
\end{equation}
Using Eq. (\ref{kka}) we obtain
\begin{equation}\label{xbsz}
4k X(z) = \rho_0^2 ( 1 + w_0 + \frac{w_a z}{1+z})^2 (1+z)^{6(w_0+w_a)} e^{-6 \frac{w_a z}{1+z}}.
\end{equation}
Following an identical procedure as before, we first plot the $F(z)$ and $X(z)$ expressions as a function of redshift  (Fig. \ref{fig3}), and then we plot the reconstructed $F(X)$ for the model. The values for $w_0$ and $w_a$ in the plots are obtained from a test of the model against observational data (see Table \ref{tI}).
Again, in Fig. \ref{fig:fig0b} we have highlighted in bold the region where the model is stable.
Notice that in contrast to Model A, in this case the stability region implies a single--valued $F(X)$ function.

\subsection{Model C}

\begin{figure}[h!]
\centering
  \begin{tabular}{@{}cc@{}}
    \includegraphics[width=.43\textwidth]{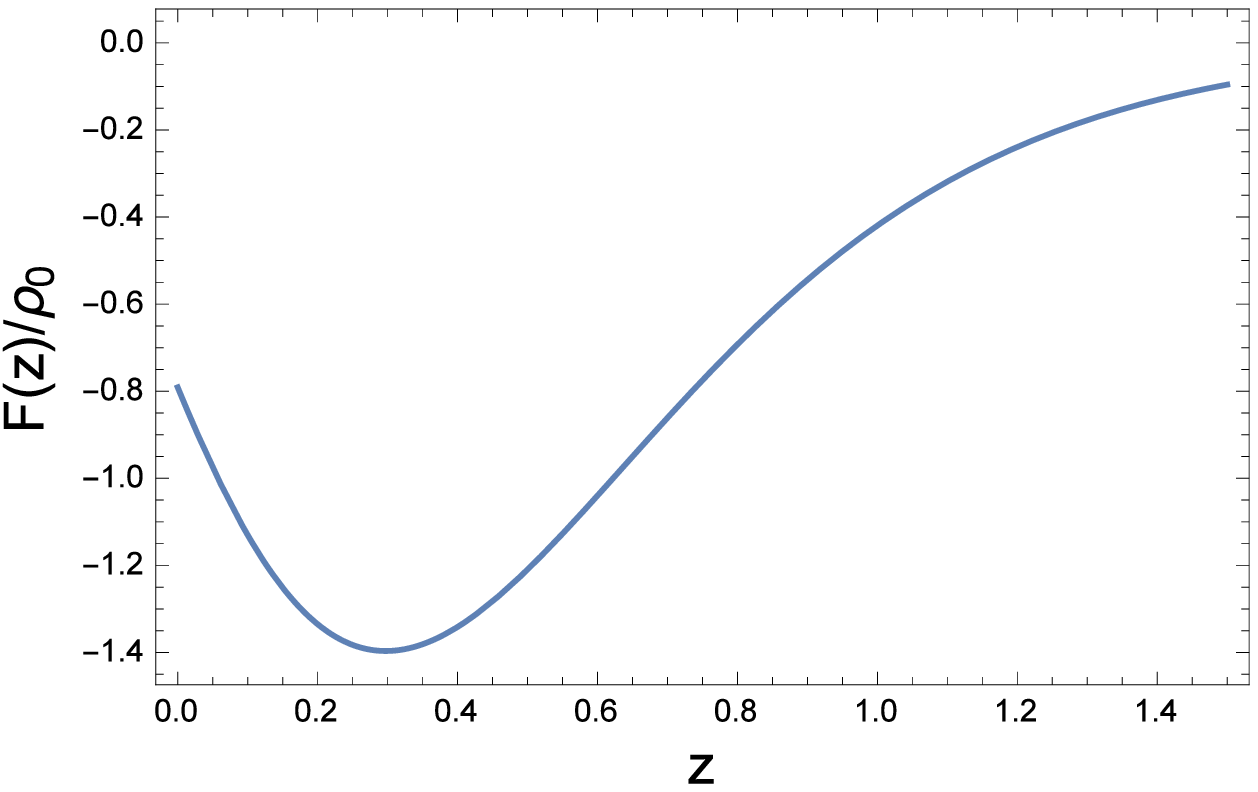} \\
    \includegraphics[width=.43\textwidth]{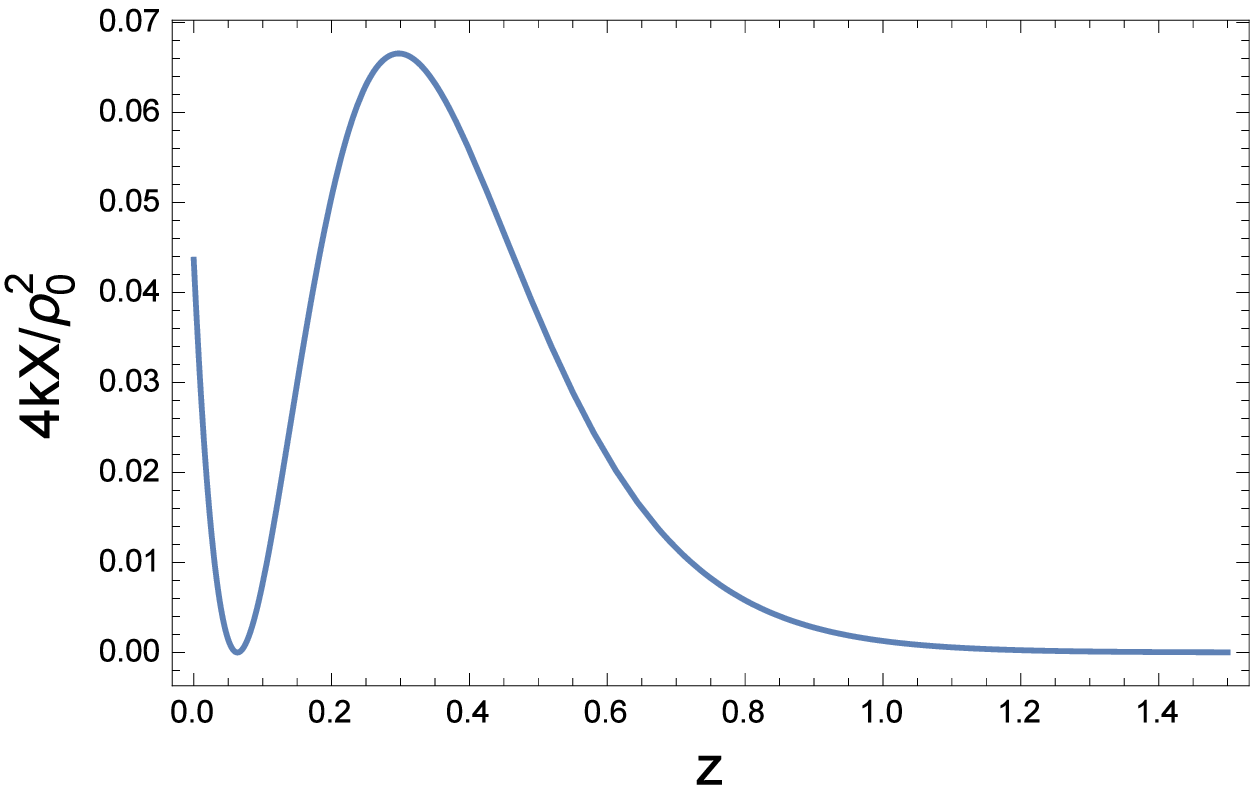}
  \end{tabular}
  \caption{Here we display $F$ and $X$ as a function of redshift
  based on Eqs. (\ref{fcsz}) and (\ref{xcsz}) for model C.
  }\label{fig5}
\end{figure}

\begin{figure}[h!]
\centering
  \begin{tabular}{@{}c@{}}
    \includegraphics[width=.43\textwidth]{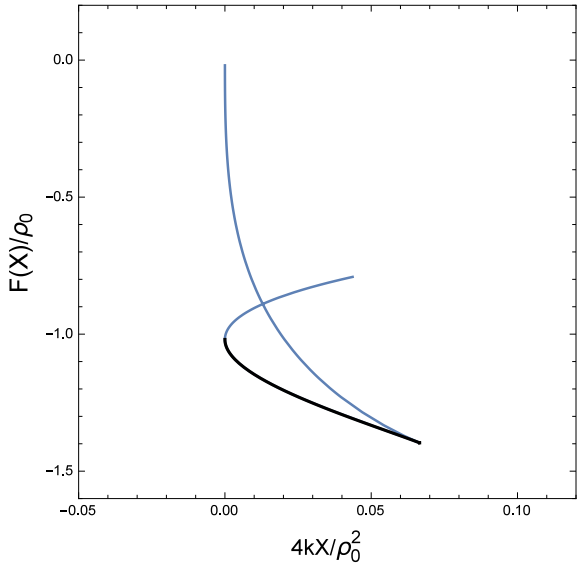}
  \end{tabular}
  \caption{Here the reconstructed $F(X)$ is shown based on the parametrization $w_0+\ln (1+z)$ with the best fit values of $w_0, w_a$ obtained from SNIa data points. The black line indicates the segment in which $c_s^2>0$.
  }\label{fig6}
\end{figure}

For model C we work with $w(z)=w_0 + w_a \ln (1+z)$. From Eq. (\ref{rhoz}) we obtain
\begin{equation}
\rho(z) = \rho_0 (1+z)^{3(1+w_0)+\frac{3}{2}w_a \ln(1+z)} .
\end{equation}
and from Eq. (\ref{ffa}) we get
\begin{equation}\label{fcsz}
F(z) = \rho_0 (w_0 + w_a \ln(1+z))(1+z)^{3(1+w_0) + \frac{3}{2}w_a \ln(1+z)} .
\end{equation}
Using Eq. (\ref{kka}) we obtain
\begin{equation}\label{xcsz}
4k X(z) = \rho_0^2 ( 1 + w_0 + w_a \ln(1+z))^2 (1+z)^{6w_0 + 3w_a \ln(1+z)} .
\end{equation}
As before, we first plot the expressions $F(z)$ and $X(z)$ as a function of redshift (Fig. \ref{fig5}), and then we plot the reconstructed $F(X)$ for the model. The best fit values for $w_0$ and $w_a$ are shown in Table \ref{tI}.
Again, in Fig. \ref{fig6} we have highlighted in bold the region where the model is stable.
Notice that the stability region implies a single--valued $F(X)$ function, as is also the case in Model B, but which does not occur in Model A.

\subsection{Model D}

\begin{figure}[h!]
\centering
  \begin{tabular}{@{}cc@{}}
    \includegraphics[width=.43\textwidth]{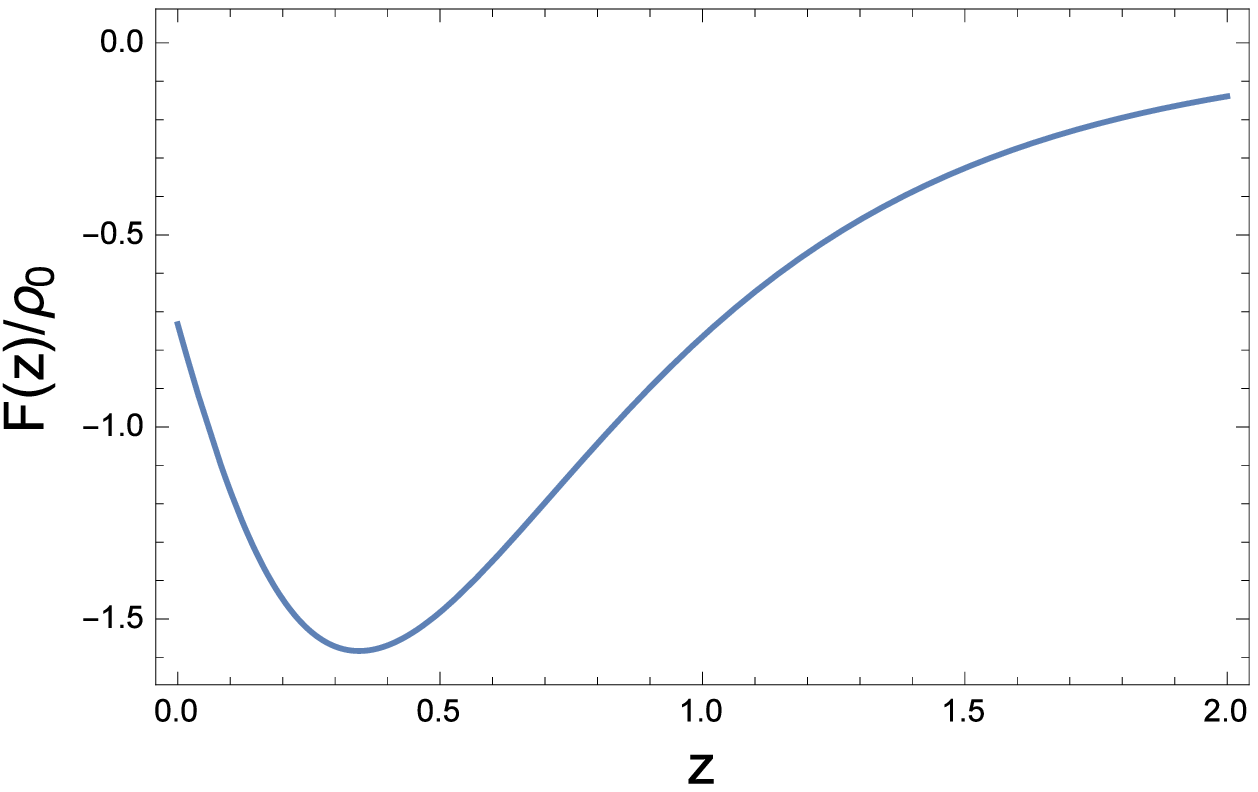} \\
    \includegraphics[width=.43\textwidth]{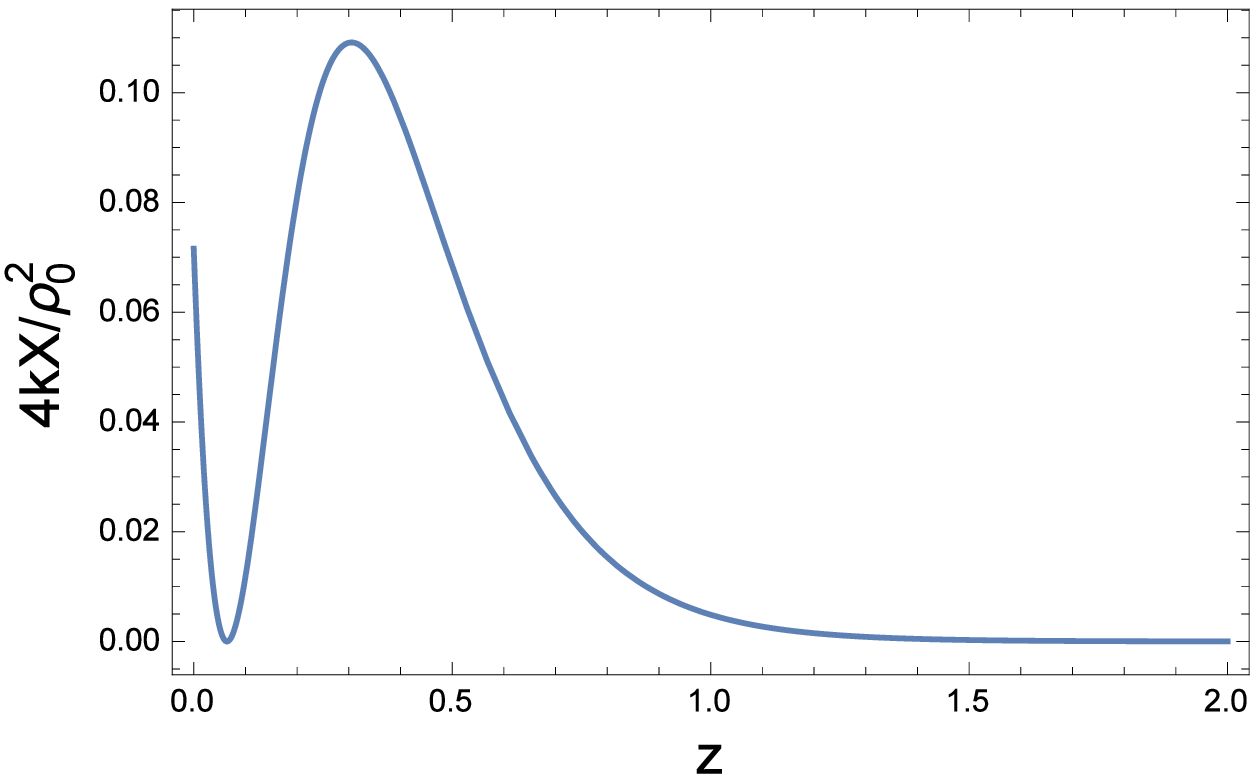}
  \end{tabular}
  \caption{Here we display $F$ and $X$ as a function of redshift derived in the text for model D.
  }\label{fig7}
\end{figure}

\begin{figure}[h!]
\centering
  \begin{tabular}{@{}c@{}}
    \includegraphics[width=.43\textwidth]{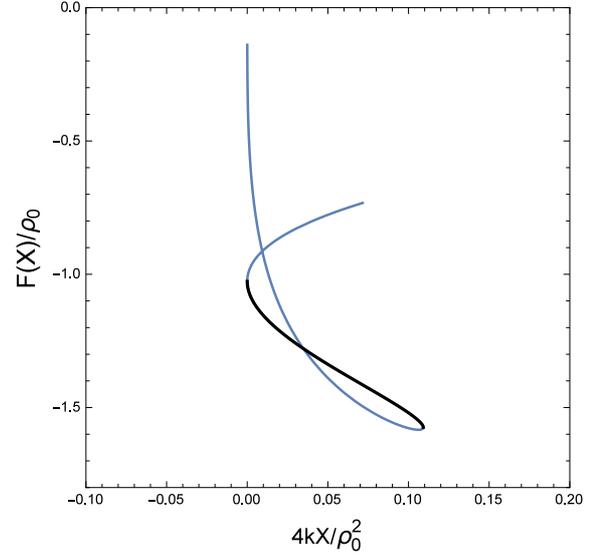}
  \end{tabular}
  \caption{Here the reconstructed $F(X)$ is shown based on the parametrization $w_0+w_a \ln(1+z/(1+z))$ with the best fit values of $w_0, w_a$ obtained from SNIa data points. The black line indicates the segment in which $c_s^2>0$.
  }\label{fig8}
\end{figure}

For model D we work with $w(z)=w_0 + \ln(1+ \frac{z}{1+z})$. From Eq. (\ref{rhoz}) we obtain
\begin{equation}
\rho(z) = \rho_0 (1+z)^{3(1+w_0)} e^{g(z)},
\end{equation}
where $g(z)$ is given by
\begin{eqnarray}
g(z)=\frac{\pi^2w_a}{4}-\frac{3w_a[\ln(1+z)]^2}{2} + 3w_a \text{Li}_2(-1-2z)+ \\ \nonumber
+ 3w_a\ln(2(1+z))\ln(1+2z).
\end{eqnarray}
Here $\text{Li}_n(z)$ is the ordinary polylogarithmic function \cite{polylog}. From Eq. (\ref{ffa}) we get $F(z)$ and using Eq. (\ref{kka}) we obtain $X(z)$. First we plot the expressions $F(z)$ and $X(z)$ as a function of redshift (Fig. \ref{fig7}), and then we plot the reconstructed $F(X)$ for the model. The best fit values of $w_0$ and $w_a$ for this model are shown in Table \ref{tI}.
Again, in Fig. \ref{fig8} we have highlighted in bold the region where the speed of sound is positive.
Notice that the stability region implies a single valued $F(X)$ function as is also the case in Model B and C, but does not occur for Model A.

\subsection{Fitting a k--essence dark energy model}

In \cite{Chim} the author proposed a family of models suitable for the description of a DE component. Certainly, a double--valued function like the one obtained from Model A is not an appropriate choice. However, the stable regions obtained from Models B, C and D are single--valued and therefore suitable for a description of this type.

In this section we test our best fit reconstructed $F(X)$ functions (cases B, C and D) against the k--essence model proposed by Chimento \cite{Chim} and defined through
\begin{equation}\label{chimmodel}
F(X)= -V_0 (1+2X^n)^{\frac{1}{2n}},
\end{equation} where $V_0$ and $n$ are constants. Clearly this is a suitable choice for our case. By inspection of Figs. (\ref{fig:fig0b}, \ref{fig6}, \ref{fig8}) for $X \simeq 0$ we get $F \simeq -1$. Extracting the data points of these figures, we perform a direct fit to the k--essence model using $V_0$ and $n$ as free parameters. The best fit values are shown in Table \ref{tII}.

\begin{table}[h!]
\begin{center}
\begin{tabular}{c|c|c}
  \hline
   & $V_0$ & $ n $ \\
   \hline
Model B  & $1.0261$ & $0.5723$  \\
Model C  & $1.0217$ & $0.5714$  \\
Model D  & $1.0221$ & $0.5490$  \\
\hline
\end{tabular}
\caption{Best fit values for models B, C and D defined in the text, using the k--essence model (\ref{chimmodel}). }\label{tII}
\end{center}
\end{table}

\begin{figure}[h]
\centering
 \begin{tabular}{@{}ccc@{}}
   \includegraphics[width=.44\textwidth]{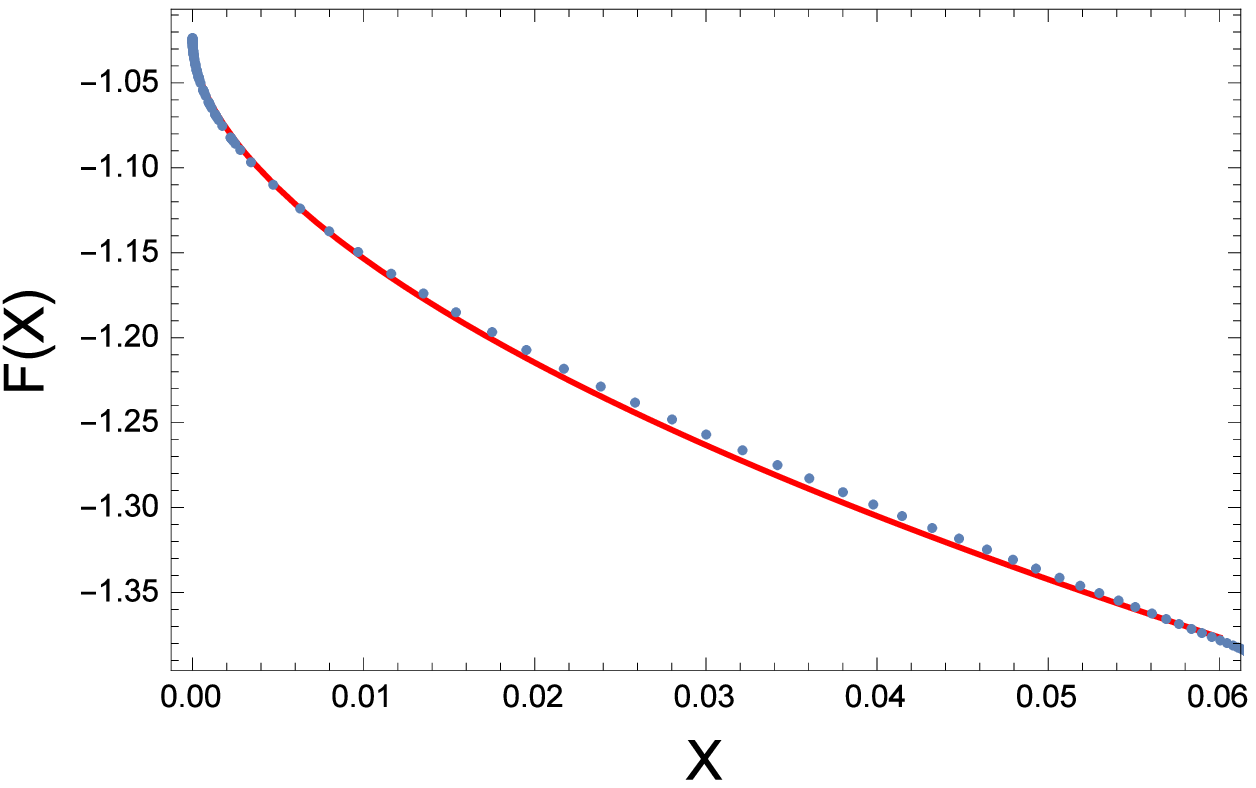} \\
   \includegraphics[width=.44\textwidth]{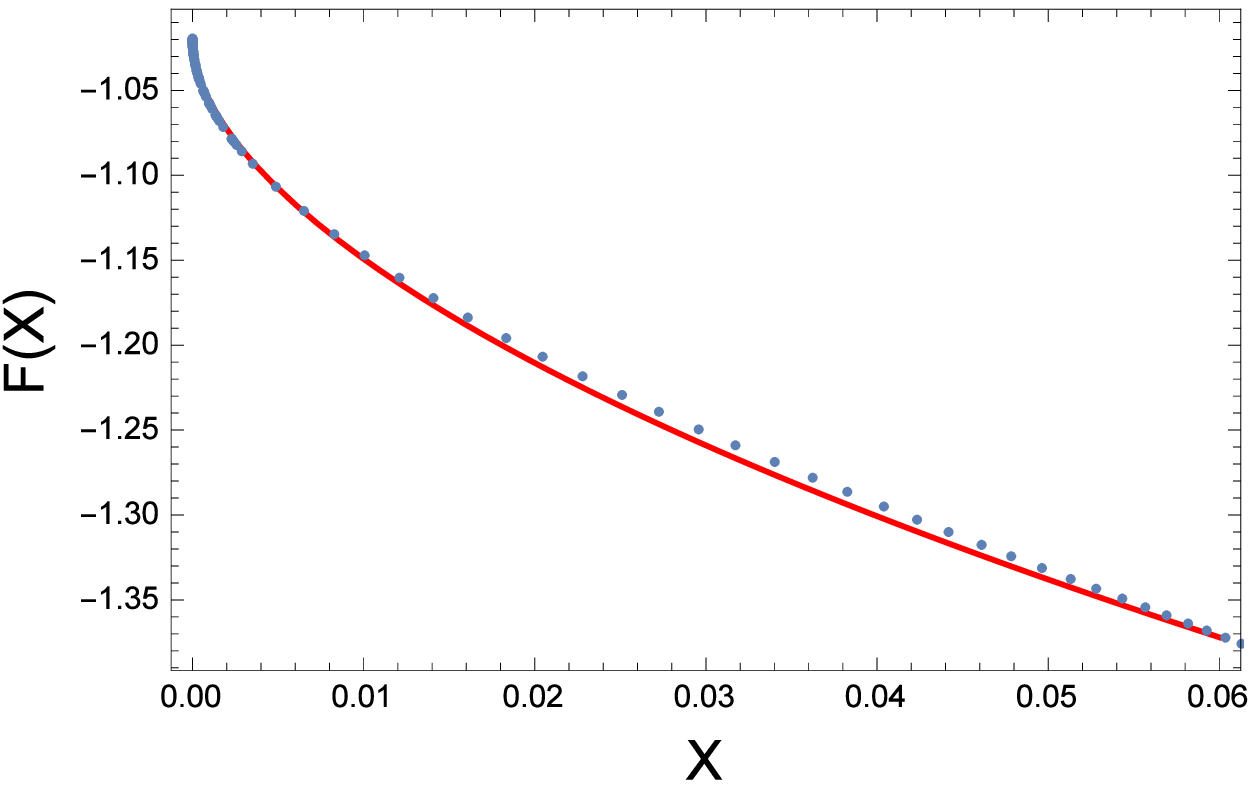} \\
   \includegraphics[width=.44\textwidth]{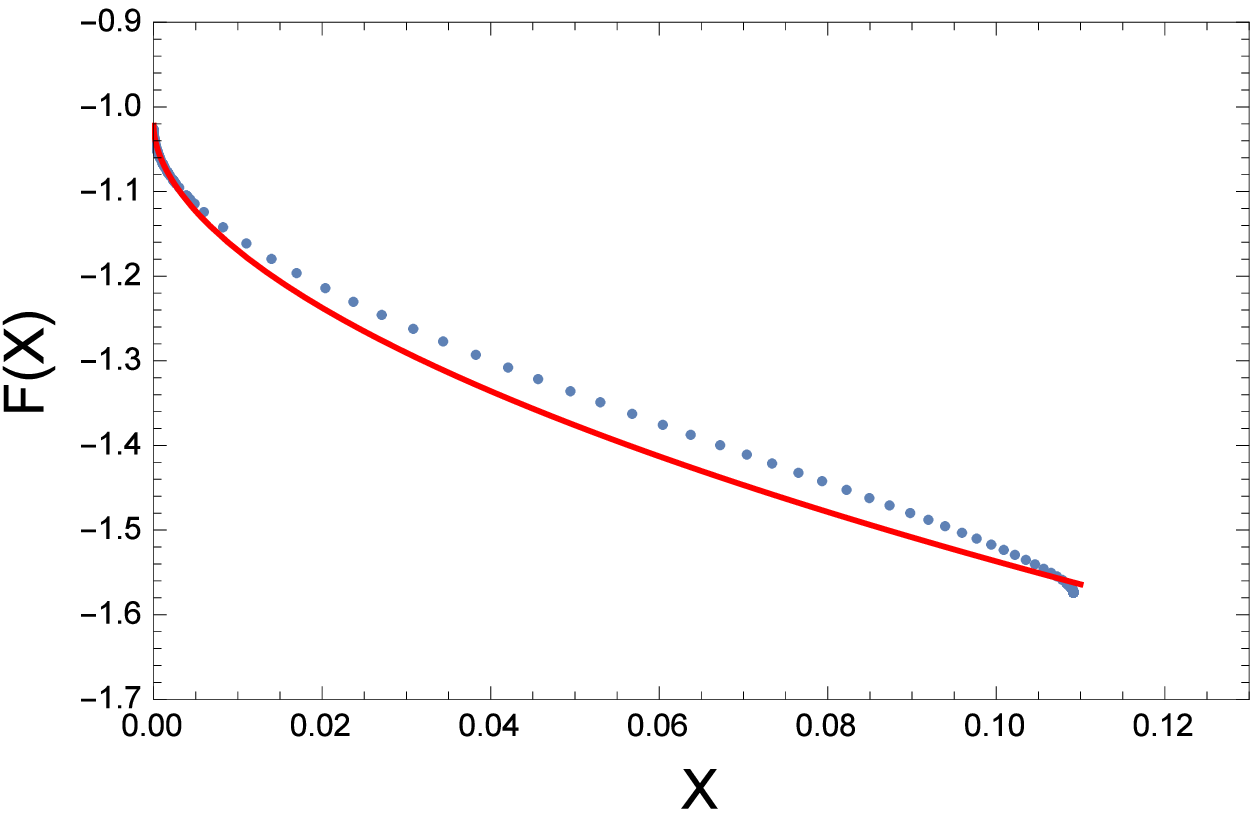}
 \end{tabular}
 \caption{TOP: Model B. MEDDLE: Model C. BOTTOM: Model D  }\label{fig:fig22}
\end{figure}

The points of the reconstructed $F(X)$ functions for models B to D are plotted together with the best fit curve of the k--essence model of Chimento (\ref{chimmodel}). These are plotted in Fig. \ref{fig:fig22}.

\section{CONCLUSIONS AND FINAL REMARKS}\label{concl}

Many phenomenological parametrizations has been proposed to describe the behavior of dark energy EoS as a function of the redshift. Our aim in this work has been to relate these phenomenological models with theoretical ones based on kinetic k--essence fields, which can be interpreted as barotropic fluid \cite{Arro} and are described by rather simple Lagrangians found in previous investigations.  We have shown the equations that can be solved numerically to reconstruct the Lagrangian density of the form $F=F(X)$ from an  initial model $\omega=\omega(a)$. The first step of this reconstruction process is to connect the Lagrangian density $F$ and the kinetic energy $X$ with the scale factor. The interval of definition of $X$ should be considered prior to the calculation of the approximation, always considering that $X>0$. Further, the equation $\frac{da}{dX}=\mathcal{H}(a,X)$ was solved to relate the kinetic energy $X$ and the Lagrangian $F$ by means of the scale factor.
Next, we compared with the observational data using the LOSS data set \cite{Ganeshalingam:2013mia}. Thus, we obtained the best values of the parameters in agreement with the observations. Also, due to the impossibility of obtaining analytical solutions, we showed the graphical solution for each model described above.

We have also fitted our reconstructed $F(X)$ using the theoretical model proposed by Chimento. Our results indicate that the observational constraints favor the model B and C instead of model D. 
As a final summary we can conclude that the phenomenological equation of states proposed for the behavior of the dark energy  may also appear  as the effective behavior of a theoretical kinetic k--essence model.    The results indicates that this equivalence can only justified in a narrow range of the redshift and for $z<1$.

\begin{acknowledgement}
N. C. acknowledges the hospitality of the Instituto de F\'isica y Astronom\'ia of the Universidad de Valpara\'iso, where part of this work was done. This research was supported by CONICYT through FONDECYT Grant No. 1140238 (NC).
\end{acknowledgement}

\end{document}